\documentclass[aps,pre,showpacs,amssymb,a4paper,twocolumn]{revtex4}%
\usepackage{amsmath}%
\usepackage{amsfonts}%
\usepackage{amssymb}%
\usepackage{graphicx}

\begin{document}

\title{Interaction regimes for oppositely charged plates with multivalent counterions}
\author{Fabien Paillusson}
\affiliation{Department of Chemistry, University of Cambridge, Lensfield Road, Cambridge, CB2 1EW, U.K.}
\author{Emmanuel Trizac}
\affiliation{Universit\'e Paris-Sud, Laboratoire de Physique Th\'eorique et Mod\`eles Statistiques, UMR CNRS 8626, 91405 Orsay, France}
\date{\today}
\begin{abstract}
Within a mean field treatment of the interaction between two oppositely charged plates in a salt free solution, the distance at which a transition from an attractive to a repulsive regime appears can be computed analytically. The mean field description however breaks down 
under strong coulombic couplings, that can be achieved at room temperature with 
multivalent counter-ions and highly charged surfaces.
Making use of the contact theorem and simple physical arguments, we propose explicit expressions for the equation of state in several situations at short distances. The possibility of Bjerrum pair
formation is addressed and is shown to have profound consequences on the interactions.
To complete the picture, we finally consider the large distance limit, from which schematic phase diagrams discriminating 
attractive from repulsive regions can be proposed.
%when they are very far from each other by first looking at the SC-PB crossover for a single plate. 
%By doing so, the interaction between the two plates can be mapped onto an interaction between two 
% effective double layers in the Poisson-Boltzmann regime which allows us to predict as well a large
%distance behaviour.
\end{abstract}
\maketitle

\section{Introduction}
$\Gamma \gg 1$ without pair contribution
Although it has been less studied than its like-charge counterpart \cite{HL00,Levin,Messina09}, 
the behaviour of two interacting oppositely charged mesoscopic bodies  
in solution is of importance in various contexts, including
colloid physics \cite{Ben-Tal,Wu,Tull,Linse,Hansen}, biochemistry related 
experiment interpretations \cite{Jonsson,Bigdeli}, drug design \cite{Morones}, 
and structural biology \cite{Jones}. The simple system of two charged plates 
with opposite uniform surface charges represents a model of choice that enables one 
to get analytical results in some limits, and furthermore, that provides a starting point to estimate the interaction energy between two colloids of various geometry \cite{DLVO,SEI}. It has been shown that opposite charge repulsion could occur within a {\it mean field} (MF) treatment \cite{Parsegian,Ohshima}. 
The physical origin of such a repulsion has been identified to be twofold: a {\it Born repulsion} due to short range polarization effects when the solvent has a dielectric constant that significantly exceeds those of the macromolecules \cite{Ben-Tal,Joanny}, and an {\it osmotic repulsion} resulting from 
the trapping of the counter-ions, that ensure electroneutrality between the unequally charged plates \cite{Andelman07,Pai09}. Within a mean-field approach for $q:q$ symmetric solutions (with $q=1$), it has been also emphasized recently that the osmotic repulsion may explain how proteins' shape 
determines their interaction with DNA \cite{Dahirel}: the essential physics of the ion 
mediated interaction between these biomolecules is well captured by a simple two plates model, which opens the way to  analytical estimates of the location and depth of the corresponding energy well.

In salt-free solutions with spherical counter-ions of size $b$, the threshold distance $D^*_{MF}=h^*_{MF}-b$ at which the electrostatic attraction is dominated by the osmotic repulsion for two plates 
bearing uniform surface charges $\sigma_1 e$ and $\sigma_2 e$ (with $\sigma_1 \sigma_2<0$), is simply given,
within mean-field, by the difference of their respective Gouy-Chapman lengths: $D^*_{MF} = |\mu_1-\mu_2|$ \cite{Pai09}. These quantities read $\mu_{1(2)}=[2 \pi q l_B |\sigma_{1(2)}|]^{-1}$, where $l_B=e^2/(4 \pi \varepsilon k_B T)$ is the Bjerrum length --about $0.7 \:\rm nm$ in water at
room temperature-- that is defined from temperature $T$ 
and solvent permittivitty $\varepsilon$. Relying on the Poisson-Boltzmann MF approximation, 
the previous result only holds provided the Coulombic coupling between counter-ions is not too large.
More specifically, this means that the two coupling parameters $\Xi_1$ and $\Xi_2$ --defined as 
$\Xi_i = 2 \pi l_B^2 q^3 |\sigma_i|$-- should both be small 
\cite{Levin,Netz01}. However, in cases of practical interest with
multi-valent counter-ions, the coupling parameter may be large; for instance, converting the
charge of double-stranded DNA into an equivalent surface charge, one finds
$\Xi\simeq 23$ in water at room temperature with di-valent ions ($q=2$) and 
$\Xi\simeq 76$ with $q=3$ \cite{Naji05}. In this paper, our goal is therefore
to study the fate of the attraction/repulsion transition for oppositely charged
interfaces, under strong coulombic coupling (large $\Xi$ limit). 

For the following discussion, it is instructive to remind the essential features
of a single strongly coupled planar double-layer, without added salt (i.e. counter-ions only
do ensure electroneutrality) \cite{Rouzina96,Shklovskii99,Netz01,Naji05,Kanduc08,Jho08,Dean09,Kanduc10,Hatlo10,Kanduc11,ST11,ST11prl}.
Irrespective of the value of $\Xi$ (from mean-field to strong coupling), 
the typical distance that counter-ions
may explore away from the charged wall is given by the Gouy length $\mu$ defined above.
At large $\Xi$ values, the counter-ions form a strongly modulated liquid (if not a true
crystal at asymptotically large $\Xi$), with a typical distance between ions 
measured by $a_\bot =\sqrt{q/\pi \sigma}$ \cite{Rouzina96}, as required by electroneutrality
($\sigma \pi a_\perp^2 \simeq q$). It therefore appears that 
$\mu \ll a_\perp$ when $\Xi\gg1$, where $a_\perp$ measures the size of the correlation 
hole around each ion.
As a consequence, the counter-ions,
that form a strongly correlated liquid parallel to the plate, 
effectively decouple in the direction perpendicular to the plate,
and the leading order profile in the strong-coupling expansion is given
by the interactions of individual counter-ions with the confining charged
interface \cite{Rouzina96,Shklovskii99,Netz01,Naji05,Kanduc08,Jho08,Dean09,Kanduc10,Hatlo10,Kanduc11,ST11,ST11prl}:
this single particle picture simply yields a leading exponential counter-ion density profile,
with a characteristic length $\mu$.
Counter-ions interactions contribute to the sub-leading terms \cite{ST11}, and will not be addressed here:
we shall restrict to the low hanging fruits of the single particle viewpoint, that provides
the dominant strong-coupling behaviour. We also stress that again for large
$\Xi$, we  not only have $\mu \ll a_\perp$ but also $a_\perp \ll q^2 l_B$.
More precisely, it is useful to keep in mind the following relations 
\begin{equation}
 \frac{a_\perp}{\mu \sqrt 2} = \frac{q^2 l_B\sqrt{2}}{a_\perp} 
 = \sqrt{\frac{q^2 l_B}{\mu}} = \sqrt{\Xi},
\end{equation}
where the numerical constants are immaterial.

\begin{figure}[h!]
 % \begin{center}
    \includegraphics[scale=0.24,keepaspectratio=true]{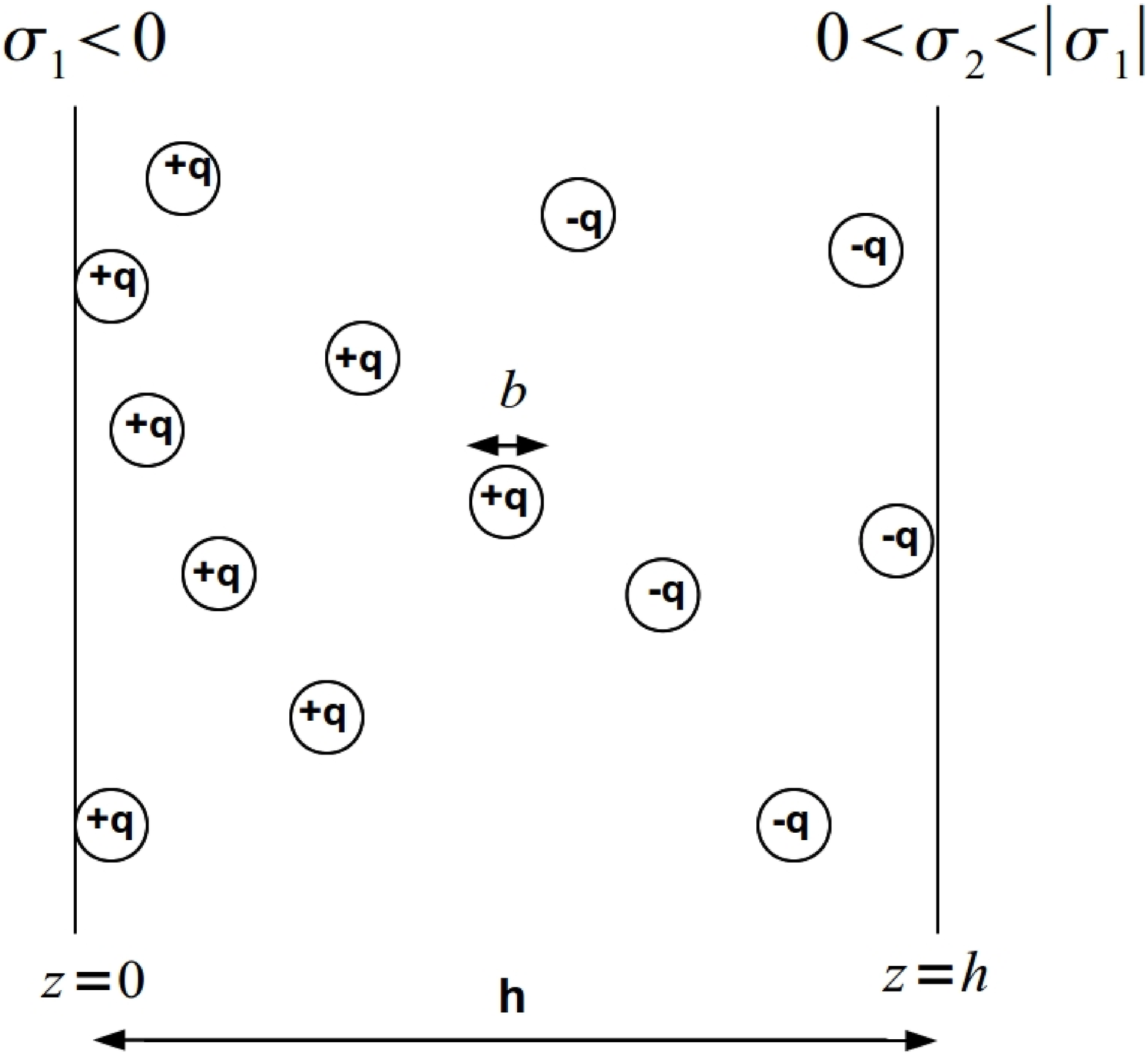}
  %\end{center}
\caption{Schematic view of the two-plates system. Micro-ions are hard spheres of
diameter $b$ with charges $+qe$ or $-q e$. 
The width of the slab between the plates is denoted by $h$, and we define $D$ as $h-b$.
}
\label{Fig1}
\end{figure}

 In the following, we shall consider two uniformly charged plates $1$ and $2$, with respective charge densities   $\sigma_1 < 0$ and $0 < \sigma_2 < |\sigma_1|$. Plate 1 is neutralized by
 counter-ions of valency $q > 0$ while $-q$ counter-ions neutralize plate 2. 
 The corresponding micro-ions remain in the gap of width $h$ between the plates to ensure global electroneutrality (see. Fig.\ref{Fig1}). Our goal is to 
characterize the strong coupling regimes, and to infer the equation of state
at short distances from the knowledge of ionic density profiles, making repeated use of the
contact value theorem  \cite{Henderson,Wennerstrom},
that will be reminded in due time. Several situations will be worked out,
depending on the formation of Bjerrum pairs $+q/-q$ between oppositely charged micro-ions. 
In addition to $\Xi_1$ and $\Xi_2$, the physics of the problem is thus ruled by another
coupling parameter, $\Gamma$, to be introduced in section \ref{ssec:pairing} and that quantifies the
tendency to form $+q/-q$ pairs. This short range study is developed in sections
\ref{ssec:without} and \ref{ssec:with}. It will be complemented by a large distance 
analysis in section \ref{ssec:large}, from which a tentative ``phase diagram'', allowing to 
discriminate repulsive from attractive regions, will be put forward.
Conclusions will be drawn in section \ref{sec:concl}, and the possible relevance 
of our approach to weak couplings will be discussed.

\section{Strong-coupling approach for oppositely charged plates} 
\label{sec:2}

\subsection{Crowding versus pairing}
\label{ssec:pairing}
\begin{figure}[h!]
 % \begin{center}
    \includegraphics[scale=0.21,keepaspectratio=true]{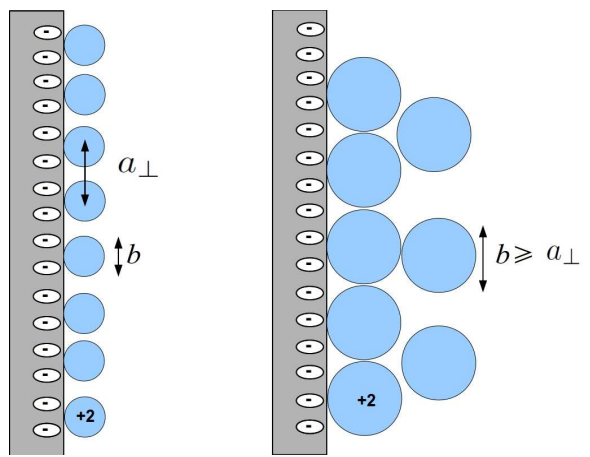}%./Fig2.eps}
  %\end{center}
\caption{{Two strongly coupled double-layers in different crowding regimes. The left hand side plot shows the uncrowded
situation where the finite ionic size does not
perturb the point-counter-ion predictions, while on the other
hand, the right hand side plot is for a case where hard core effects lead 
to crowding, with bi or multi-layers of counter-ions in the vicinity of the plate. 
For this example to be meaningful, we set $q=+2$ for the valency of the counterions.}}
\label{fig:pack}
\end{figure}
Whereas previous works pertaining to the strong-coupling limit have been
mostly performed in the limit of point counter-ions, 
it is possible to transpose in some cases the results to the case of finite size ions,
essentially by taking $b/2$ (the ionic radius), as the ion-plate distance of closest 
approach. For a single plate, 
the density profiles in the two cases are therefore identical,
up to a coordinate shift $z \to z-b/2$, where $z$ denotes the distance to the plate. 
Likewise, in the two plates problem, the plate-plate distance of closest approach
is $b$. More precisely, the $b=0$ and $b\neq 0$ cases coincide provided
packing effects are negligible (see Fig. \ref{fig:pack}-left) while increasing ionic
size $b$ necessarily leads to a situation where $b$ becomes of the order
of $a_\perp$, so that the double-layer can no longer accommodate a mono-layer
of counter-ions (see Fig. \ref{fig:pack}-right). 
Understanding the behaviour of strongly coupled and crowded double-layers
is an open problem that lies beyond the scope of the present work, so that we will restrict to the 
cases where $b<a_\perp$, i.e. to not too big micro-ions. This
requirement should be enforced for both plates: $b< a_{\bot}^{(1),(2)}$.
\begin{figure}[h!]
 % \begin{center}
    \includegraphics[scale=0.21,keepaspectratio=true]{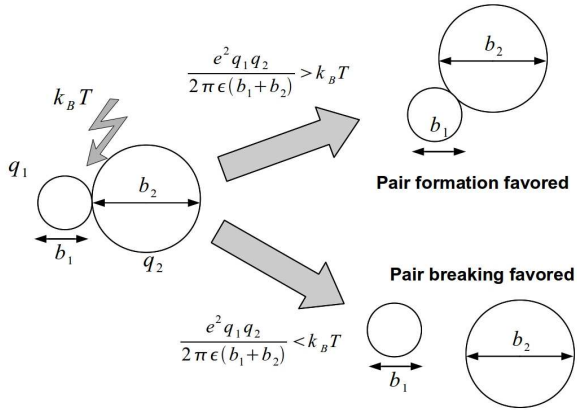}
  %\end{center}
\caption{{ Pair breaking/formation mechanism. Starting from a contact configuration for two ions of opposite charges $q_1 e$ and $-q_2 e$ on the left hand side, the tendency to remain in this configuration, 
or, on the contrary, to break the pair, is given by comparing the electrostatic loss to the thermal energy.}}
\label{Fig2}
\end{figure}

In addition to crowding, micro-ion pairing may take place in the two plates
problem \cite{Levin,Roij}, see Fig.\ref{Fig2}. The tendency for $+q$ and $-q$ micro-ions
to form neutral pairs at $T \neq 0$ is quantified by the ratio between the 
direct electrostatic interaction at close contact and $k_B T$ i.e. $\Gamma=q^2 l_B/b$. 
Interestingly, keeping in mind the ``no-crowding'' condition sketched above
($b< a_{\bot}^{(1),(2)}$), we get the inequality $2 \Gamma^2 > \Xi$ (where $\Xi=\rm max(\Xi_1,\Xi_2)=\Xi_1$), so that the possible values $\Gamma$ can take are bounded from 
below by $\sqrt{\Xi/2}$. Consequently, strongly coupled uncrowded double-layers
lead to the important formation of Bjerrum pairs (large $\Gamma$). 
We nevertheless start
by considering the rather narrow region where $3 \lessapprox \sqrt{\Xi/2}<\Gamma <10$ in which pair formation can be neglected. The above constraint translates into $20<\Xi<200$,
where $\Xi$ is large enough to allow for a strong coupling analysis in due form
to unveil the main features. A more quantitative description 
presumably requires, especially at the smaller $\Xi$-values involved, an intermediate approach 
interpolating between the mean-field and strong-coupling limits \cite{Burak,Chen06,Santangelo06,Buyukdagli10}.

\subsection{Small separation distances without pair formation}
\label{ssec:without}

\begin{figure}[h]
   \begin{center}
   \includegraphics[scale=0.24,keepaspectratio=true]{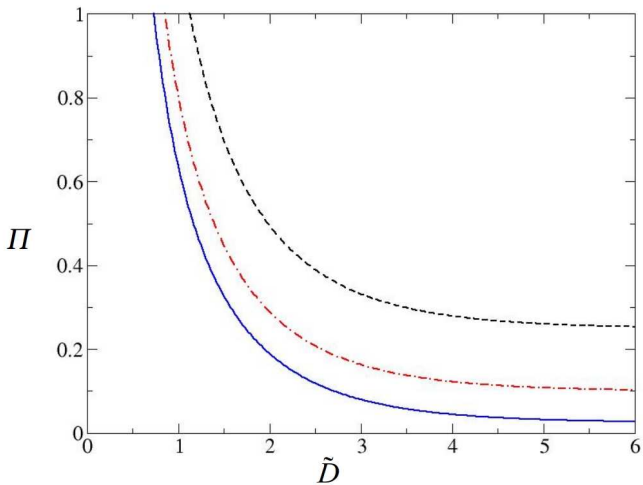}
   \end{center}
\caption{{ Plot of the reduced pressure following from the short-distance equation of
state (\ref{eq6}), as a function of rescaled distance $\widetilde D$, for several 
values of $\zeta=\sigma_2/|\sigma_1|$: $\zeta=0.25$ (dashed line), $\zeta=0.1$ (dot-dashed line) and $\zeta=0.025$ (solid line). The short-distance requirement $D<a_\perp^{(1)}$
(the so-called Rouzina-Bloomfield criterion \cite{Rouzina96}),
translates into $\widetilde D < \Xi^{1/2}$. 
}}
\label{fig:repulsions}
\end{figure}

We define in the subsequent analysis $D$ as the shifted distance between the plates~:
$D=h-b$. The first situation addressed is that where Bjerrum pair formation can be neglected,
which is the assumption made in \cite{Kanduc08}. Under strong coupling, 
if $D< a_{\bot}^{(1)}$ (which implies that $D<a_{\bot}^{(2)}$ since $\sigma_2 < |\sigma_1|$),  
the single particle picture where each micro-ion only interacts with both plates and not with its
fellow micro-ions is valid. A tagged micro-ion feels an electric field 
$-2 \pi (|\sigma_1|+\sigma_2) \hat z/\varepsilon$
where $\hat z$ is a unit vector along the $z$ direction, 
hence a linear potential in $z$. 
The corresponding number densities $n_+(z)$ and $n_-(z)$ for both $+q$ and $-q$ species
follow then a simple Boltzmann law:
\begin{equation}
 n_{\pm}(z)=n^{(0)}_{\pm} e^{ \mp \tilde{z} }  \label{eq1}
\end{equation}where $n^{(0)}_+$ and $n_-^{(0)}$ are two normalization constants, 
and where we introduced the reduced distance to plate $1$, 
\begin{equation}
\tilde{z}=z/\lambda \quad \hbox{with} \quad \frac{1}{\lambda}= \frac{1}{\mu_1}+\frac{1}{\mu_2}.  
\end{equation}
The two factors $n^{(0)}_{\pm}$ can be determined from the electroneutrality conditions:
\begin{eqnarray}
 && q\int_{b/2}^{h-b/2} dz\:n^{(0)}_+ = |\sigma_1| \label{eq2} \\
 && q\int_{b/2}^{h-b/2} dz\:n^{(0)}_- = \sigma_2,
 \label{eq3}
\end{eqnarray}
so that
\begin{eqnarray}
&& n_+(\tilde{z})=\frac{ |\sigma_1| e^{\tilde{b}/2-\tilde{z}}}{q \lambda(1-e^{\tilde{b}-\tilde{h})}} \label{eq4} \\
&& n_-(\tilde{z})=\frac{ \sigma_2 e^{-\tilde{b}/2+\tilde{z}}}{q \lambda(e^{-\tilde{b}+\tilde{h}}-1)}. \label{eq5}
\end{eqnarray}
The expression for the reduced pressure
\begin{equation}
\Pi \, = \, 2\pi \,l_B \,q^2 \,\mu_1^2 \,\beta P \, =\, \frac{\beta P}{2 \pi \,l_B \,\sigma_1^2}
\end{equation}
directly follows from the contact value theorem, which yields the pressure $P$ in the form \cite{Henderson,Wennerstrom}:
\begin{eqnarray}
 P &=& n_+\left(\frac{b}{2}\right) + n_-\left(\frac{b}{2}\right) \,-\, 2 \pi  l_B \sigma_1^2 
 \label{eq:Pplate1}\\
   &=& n_+\left(h-\frac{b}{2}\right) + n_-\left(h-\frac{b}{2}\right) \,-\, 2 \pi  l_B \sigma_2^2.
\end{eqnarray}
Consequently, we have
\begin{equation}
  \Pi(\widetilde{D}) \,= \, \zeta \coth \left(\frac{\widetilde{D}}{2}\right)+\frac{1}{2}(1+\zeta^2)
 \left[\coth\left( \frac{\widetilde{D}}{2}\right) -1\right] 
 \label{eq6}
\end{equation}
where we introduced the charge ratio $\zeta=\sigma_2/|\sigma_1|$. 
Equation \eqref{eq6} is independent of the plate (1 or 2) 
where the contact theorem is applied, which provides a consistency test for the argument. 
In other words, the pressure $P$ is invariant under the change $\zeta \to 1/\zeta$,
so that the reduced pressure should change according to $\Pi \to \Pi \zeta^{-2}$
when $\zeta \to \zeta^{-1}$.
This property can be checked directly on Eq. (\ref{eq6}).
More importantly, expression \eqref{eq6} is positive for $D>0$,
for all values of the charge ratio, see Fig. \ref{fig:repulsions}. Therefore, 
the interaction between two oppositely charged plates is always repulsive at short distances 
in this regime. The physical mechanism behind this repulsive
behaviour is the following. Compared to the large distance limit
where $n_+(b/2) \simeq 2 \pi  l_B \sigma_1^2 $, as follows from 
Eq. (\ref{eq:Pplate1}) and the fact that both the pressure and $n_-(b/2)$ vanish,
bringing the plates at short distances where $D<a_\perp^{(1)}$
enhances the electric field felt by $+q$ micro-ions, which has the
result to increase their density at contact with plate 1. Invoking
again the contact theorem (\ref{eq:Pplate1}), the consequence is that $P>0$. 
The interactions between $+q$ and $-q$ micro-ions could counter-balance this
effect, but these interactions have been discarded here, with the neglect of Bjerrum
pair formation. We will see below that $+q$/$-q$ interactions,
when relevant, completely change the phenomenology.

Three remarks are in order here. 1) We see that the reason for 
observing repulsive behaviour (enhanced electric field acting on a micro-ion within the
single particle picture) is the same as that leading to attraction in the
like-charged case (decreased electric field, with a corresponding decrease
of micro-ionic density at contact; this effect is most pronounced in the symmetric
case where $\sigma_1=\sigma_2$, for which the electric field vanishes
and the micro-ion densities is uniform in the $z$ direction). 2) The possibility
of attraction under strong coupling reported in \cite{Kanduc08}
for oppositely charged plates stems from the fact that only one type
of micro-ion was considered in \cite{Kanduc08}. This results in a smaller
amount of counter-ions (per unit surface), compared to that which is necessary
to neutralize the isolated plate, with a concomitant decrease of contact
ionic density, which opens the way to a possible attraction.
The physical situation considered in \cite{Kanduc08}
thus differs from the one under study here.
3) Our finding $P>0$ relies on the condition $D<a_\perp^{(1)}$.
At large distances, we should have $P<0$ since a mean-field scenario
is then expected to prevail \cite{Shklovskii99,Burak,Boroudjerdi05,dosSantos}. We will come back to this point
in section \ref{ssec:large}.

\subsection{Small separation distances with pair formation}
\label{ssec:with}
We now turn to the case where $1 \ll \sqrt{\Xi/2} < \Gamma$, with a strong tendency 
for two micro-ions to form a neutral pair.
While pair formation is unlikely as long as the two condensed micro-ion 
layers from each plate do not overlap, it turns important at smaller separations.
By ``pair formation'', we loosely refer here to the more or less complex structures, or aggregates,
that may form from the association of several of individual pairs.
Pairs may indeed exist in the form of well defined entities, but may also self assemble
into chains, see e.g.  Ref. \cite{WL93}, or into more complex 
structures (regular or empty crystals) uncovered in a related context in Ref. \cite{AML10}.
The corresponding aggregates are electrically neutral, with 
number of Bjerrum pairs involved per unit area
limited by the less abundant species of micro-ion, i.e. the counter-ions of the plate $2$.
Therefore, the aggregate surface density is bounded from above by $\sigma_2/q$. 
These aggregates  coexist with a strongly correlated
Wigner-like crystal made up of the remaining majority species.
In this work, we did not attempt at a precise evaluation of the aggregate,
or ``pairs'' contribution $P_{agg}$ to the total interplate pressure $P$, but instead, we considered
two limiting cases, where we bound $P_{agg}$ from below by 0 (see section \ref{ssec:nopairs}),
and from above by $kT \sigma_2/(q D)$ (see section \ref{ssec:osmowith}).
The latter bound corresponds to a density of aggregates, that are neutral entities,
equal to $\sigma_2/(qD)$, that is, to the maximum mean density of possible pairs. Any 
self-assembly of the pairs in a more complex architecture leads to a decrease of that
density. 
We now investigate separately these two limiting cases.

%Two limiting cases will be considered below,
%depending on the inclusion of the pairs contribution to the osmotic pressure.

\subsubsection{Without the osmotic contribution from the pairs}
\label{ssec:nopairs}
 
With only positive counter-ions in the system, the typical lateral distance becomes $A_{\bot}=\sqrt{q/\pi(|\sigma_1|-\sigma_2)}$ and 
%the SC regime is well defined if $q^2 l_B/A_{\bot} >1$ i.e. if 
%$(\Xi_1-\Xi_2) \gg 1$. We will see below that this 
%criterion can be reformulated into $\Xi_1 \gg \Xi_2 \gg 1.$
%If that is the case, then 
for separation distances $D$ less than $A_{\bot}$, the single particle picture holds, and yields the micro-ionic density $n_+(z)$ 
in a form similar to \eqref{eq1}:
\begin{equation}
n_+(z)=n^{(1)}_+ e^{- \tilde{z}} \label{eq7}
\end{equation}
where $n^{(1)}_+$ is a positive constant.
As for $n^{(0)}_+$, the $n^{(1)}_+$ prefactor can be determined using the electroneutrality condition:
\begin{equation}
 q\int_{b/2}^{h-b/2} dz \:n_+(z)=|\sigma_1|-\sigma_2 \label{eq8}
\end{equation}
and the ion density then reads:
\begin{equation}
 n_+(z)= \frac{ |\sigma_1|-\sigma_2}{ q\lambda (1-e^{\tilde{b}-\tilde{h}})} e^{\tilde{b}/2-\tilde{z}}. 
 \label{eq9}
\end{equation}

In a first step, we do not consider the contribution of Bjerrum pairs to the total pressure
(we therefore bound from below the term $P_{agg}$ by 0). In doing so, 
the pressure at a given reduced separation distance $\widetilde{D}$ can again be found by means of the contact value theorem, with only one species of micro-ions: 
$\beta P = n_+\left(b/2\right)  - 2 \pi  l_B \sigma_1^2 $, so that
\begin{equation}
 \Pi(\widetilde{D}) = -\frac{1}{2}(1+\zeta^2)+\frac{1}{2}(1-\zeta^2)\coth(\frac{\widetilde{D}}{2}) 
 \label{eq10}
\end{equation}
We recover the same expression as in \cite{Kanduc08}, from a ``mechanical'' (contact theorem)
instead of ``energy'' route.
Unlike Eq.\eqref{eq6}, Eq.\eqref{eq10} does not have a definite sign and as a consequence, 
the interaction is attractive at large distances: there exists a threshold
\begin{equation}
 D^*_0 = 2\lambda\ln( |\sigma_1|/\sigma_2) 
 \label{eq11}
\end{equation}
below which the interaction becomes repulsive ($\Pi>0$), see Fig. \ref{Fig3}. We note from Eq. \eqref{eq11} 
that attraction prevails until $D \rightarrow 0$ when $\sigma_2 \rightarrow |\sigma_1|$ and we add
that as long as micro-ions (even in small amount) remain between the plates as required by electroneutrality, the corresponding entropy cost for confinement makes the pressure
diverge (hence positivity) for $D\to 0$. Only for $\zeta=1$, i.e. $\sigma_2 = -\sigma_1$
would micro-ion total density vanish, which leaves two oppositely charged plates
interacting without any screening. In that specific case, 
the interaction is obviously attractive until close contact $h=0$. 
On the other hand, in the large  $\widetilde{D}$ limit --with nevertheless the $D<A_\perp$ requirement 
enforced--, and for any value of $\zeta$,
one can immediately find the pressure from the contact theorem applied at plate
2: for large $\widetilde{D}$, the positive micro-ions are expelled from the vicinity of the
positive plate, which means that the contact density $n_+(h-d/2)$ vanishes and that
$\beta P \to - 2 \pi l_B \sigma_2^2$. In terms of rescaled pressure, we then
have $\Pi \to -\zeta^2$, which is indeed observed in Fig. \ref{Fig3}.

\subsubsection{With the osmotic contribution from the pairs}
\label{ssec:osmowith}

\begin{figure}[h]
   \begin{center}
   \includegraphics[scale=0.24,keepaspectratio=true]{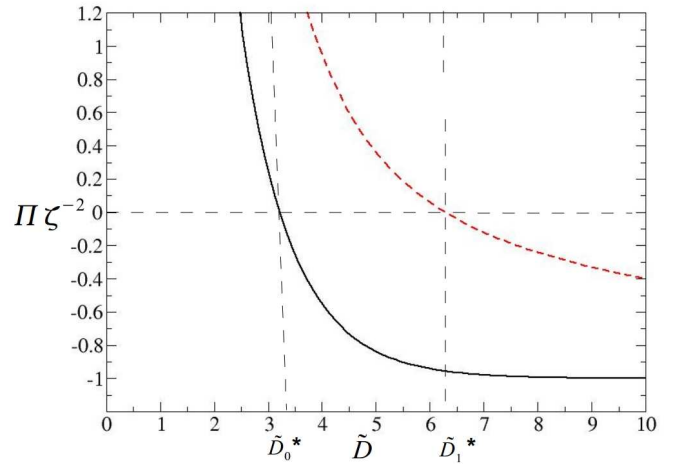}
   \end{center}
\caption{{ Pressure curves from Eqs. \eqref{eq10} (solid line) and \eqref{eq12} (dashed line) for $\zeta=0.2$.
%For the SPP to hold we took more specifically $\Xi_2=2\pi l_B q^3\sigma_2 \approx 5$. 
}}
 \label{Fig3}
 \end{figure}

\begin{figure}[h]
   \begin{center}
   \includegraphics[scale=0.25,keepaspectratio=true]{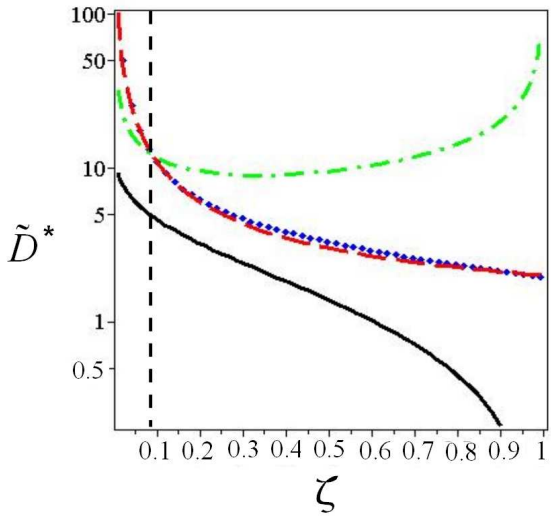}
   \end{center}
\caption{Reduced threshold distance discriminating (strong-coupling) attraction
from repulsion, as a function of $\zeta$, in semi-log scale. The solid line shows $\widetilde{D}_0^*$
obtained from \eqref{eq10}, the dotted line is for $\widetilde{D}_1^*$, the exact root of Eq. (\ref{eq12}),
and the dashed line displays the approximation \eqref{eq14}. The upper dot-dashed line
shows $\tilde A_\perp$, in the particular case $\Xi_2=25$: 
our approach is meaningful for $\widetilde D < \tilde A_\perp$ only, which translates into
$\zeta > 1/(2 \Xi_2+1)$, see text. Here, $1/(2 \Xi_2+1)\simeq 0.02$.}
\label{fig:D0D1}
\end{figure}

We now include the pairs contribution to the equation of state, through the upper
bound $\sigma_2/(qD)$ alluded to above. 
We then get
\begin{equation}
  \Pi = \frac{(1+\zeta)\zeta}{\widetilde{D}}-\frac{1}{2}(1+\zeta^2)+\frac{1}{2}(1-\zeta^2)\coth(\frac{\widetilde{D}}{2}) 
  \label{eq12}
\end{equation}
Clearly, compared to expression \eqref{eq10}, the effect of this osmotic contribution is to increase the threshold value where repulsion ($\Pi>0$) can be observed. The two limiting behaviours,
Eq. (\ref{eq10}) and Eq. (\ref{eq12}),
are sketched in Fig. \ref{Fig3}. The corresponding values of the thresholds $\widetilde{D}_0^*$ and $\widetilde{D}_1^*$ are indicated. These two quantities are plotted in Fig. \ref{fig:D0D1}
as a function of charge asymmetry,
together with the analytical estimation of $\widetilde{D}_1^*$ obtained as follows.
If $\widetilde{D}$ is large enough,  Eq. \eqref{eq12} simplifies to:
\begin{equation}
 \Pi \simeq -\zeta^2+\frac{(1+\zeta)\zeta}{\widetilde{D}}.
 \label{eq13}
\end{equation}
This expression can now exhibit a repulsive behaviour below $\widetilde{D}=(1+\zeta)/\zeta$ i.e.
\begin{equation}
D^*_1 \simeq \mu_2 
\label{eq14}
\end{equation}
It can be seen that this approximation (dashed line) is in fair agreement with the root of Eq.
(\ref{eq12}) found numerically (dotted line), in the whole available range.

The single particle picture invoked here relies on two assumptions. 
First, $\Gamma$ should be large, to have pair formation.
%Second, the coupling should be strong enough, i.e. $\Xi_1\gg \Xi_2+1$. 
%At variance with our analysis in section \ref{ssec:without}, we do not need here 
%to have $\Xi_2 \gg 1$, but only a large enough $\Xi_1$. The reason is that ``plate 2 micro-ions''
%are here ``evacuated'' in the pairs. 
Second, the (shifted) distance $D$ between the plates should not exceed $A_{\bot}$,
see the upper dot-dashed line in Fig. \ref{fig:D0D1}.
Making use of approximation (\ref{eq14}), this means $1/(2 \Xi_2+1) < \zeta$. 
For smaller values of $\zeta$, the analysis is significantly more
complex (loss of the single particle view-point). 
Hence, if $\zeta > 1/(2 \Xi_2+1)$ the interaction is repulsive at short distances and then turns 
attractive at intermediate distances,
while if $ \zeta < 1/(2 \Xi_2+1)$, Eq.\eqref{eq13} does not lead to any transition 
between repulsion and attraction, and is always repulsive in its range of validity,
as was the case for Eq.\eqref{eq6}. Of course, for large $\Xi_2$, the threshold
$(2 \Xi_2+1)^{-1}$ is small, so that extremely asymmetric cases only (very
low $\zeta$ are not covered by our analysis.

%Note also that the two curves in Fig.\ref{Fig3} correspond to extreme situations where on one hand the osmotic contribution from the pairs can be completely neglected while on another hand it is considered at its maximal value. In reality, the physics of the system can be such that only a partial amount of created pairs will contribute to the pressure. For instance one could imagine dipole-dipole interactions leading to chains formation in the slit and then to a lower resulting osmotic contribution. In any case the corresponding pressure curve would lie in between the two curves displayed in Fig.\ref{Fig3} if dipole-plate interactions can be neglected.

\subsection{Large separation distances}
\label{ssec:large}

Our analysis has so far been restricted to short distance expansions. We are now interested in  
large distance asymptotics and in attempting to match the short and large distance behaviours. In doing
so, we will discuss qualitatively an attraction/repulsion transition of an ``effective'' mean-field
type, which leads to reentrant attraction as the distance between the two plates 
is varied from infinity to close contact. 

\subsubsection{Crossover between strong-coupling and mean-field regions for one plate}

\begin{figure}[h!]
   \begin{center}
    \includegraphics[scale=0.20,keepaspectratio=true]{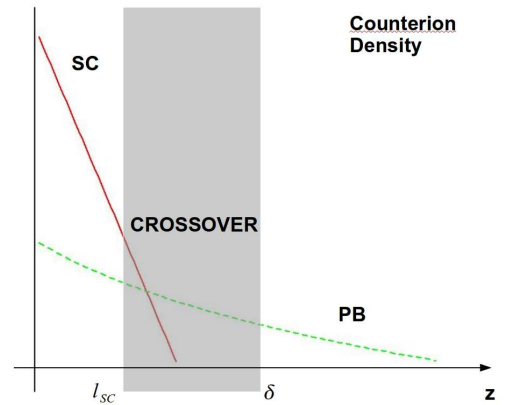}
 \end{center}
 \caption{Crossover scheme: following Ref. \cite{dosSantos},
 schematic representation in semi-log scale of the crossover 
between strong coupling (SC) and mean-field regimes for the counter-ion density in the vicinity
of a single highly charged plate. The solid curve represents the exponential decay expected close to the plate while the dashed curve stands for the algebraic decay expected 
at large distances where the Poisson-Boltzmann theory (PB) should hold. The shaded region corresponds to the crossover between these two regimes: $l_{sc}$ is the distance beyond which the exponential decay no
longer holds and $\delta$ is the distance to the plate beyond which the PB profile is expected to be valid.}
 \label{Fig4}
 \end{figure}

We will assume first that a given strongly-coupled plate (having thus a large
$\Xi_i$), can be effectively described by mean-field theory, at sufficiently
large distances $z$. This ``common wisdom'' stems on the remark that for large 
$z$, the typical distance between counter-ions becomes large, which leads
to a low coupling regime \cite{Shklovskii99,Burak,Boroudjerdi05,dosSantos}.
It should be emphasized though that the above point of view, that predicts
a large $z$ density decay in $1/z^2$, is incorrect in two dimensions, 
as shown in a recent work \cite{ST11}. The present
study pertains to three dimensional systems, so that we nevertheless expect for a single 
plate the crossover scenario discussed in Ref. \cite{dosSantos},
and summarized in Fig. \ref{Fig4}. In essence, the density is expected to decrease 
exponentially at short distances, and algebraically at large distances: beyond 
a distance $\delta$ from the plate, the counter-ion density $n$ is simply given by the solution of the nonlinear Poisson-Boltzmann (PB) equation:
\begin{equation}
 n(\mathfrak{z})=\frac{1}{2 \pi l_B q^2 (\mathfrak{z}+\mu^{\rm eff})^2} \label{eq15}
\end{equation}where $\mathfrak{z} \equiv z-b/2$ and where $\mu^{\rm eff}$ is an effective Gouy-Chapman length characterizing this long range behaviour. Following \cite{dosSantos}, one can match 
the two regimes by assuming that the condensed counter-ion layer forms a 2D One Component Plasma and by applying a mean-field approximation for the dilute layer. Equating the two corresponding chemical potentials yields
\begin{equation}
 n(\delta)=n_{sc}e^{-\beta |\varepsilon_c|} \label{eq16}
\end{equation}
where $n_{sc}$ is a $\delta$-related average density in the condensed layer and  $\beta\varepsilon_c(\Xi)\approx -1.56 \sqrt{\Xi}$ is the contribution to the 2D one component
plasma chemical potential that stems from the correlations between the counter-ions \cite{Shklovskii99,Levin}. 
Extrapolating the validity of Eq.\eqref{eq15} to $\delta \rightarrow b/2$ and assuming that in such a situation, $n_{sc}$ is well approximated by the average density over the characteristic length $l_{sc}$, 
we arrive at \cite{dosSantos}:
\begin{equation}
 n(\mathfrak{z}=0)=\frac{ |\sigma|}{q l_{sc}} \left[ 1- e^{-l_{sc}/\mu} \right]e^{-1.56 \sqrt{\Xi}}. \label{eq17}
\end{equation}
%\begin{equation}
 %n_c(\mathfrak{z}=0)=\frac{2 \pi l_B \sigma^2}{3.701}e^{-1.56\beta \sqrt{\Xi}} \label{eq17}
%\end{equation}
Equation \eqref{eq17} is nothing but the density that the mean-field profile, 
valid at large distances from the plate, would have if extrapolated at $\mathfrak{z}=0$, and is 
therefore not the real density at the plate. However, invoking Eq.\eqref{eq15}, it allows one to estimate the effective Gouy-Chapman length $\mu^{\rm eff}$ corresponding to the charged plate {\it dressed} by a condensed counter-ion layer, which will prove useful in the following. For the subsequent quantitative discussion, we shall
take the value $l_{sc} \approx 3.6 \mu$, already used in \cite{dosSantos}.

\subsubsection{Application to the two plates problem}

\begin{figure}[h!]
   \begin{center}
    \includegraphics[scale=0.25,keepaspectratio=true]{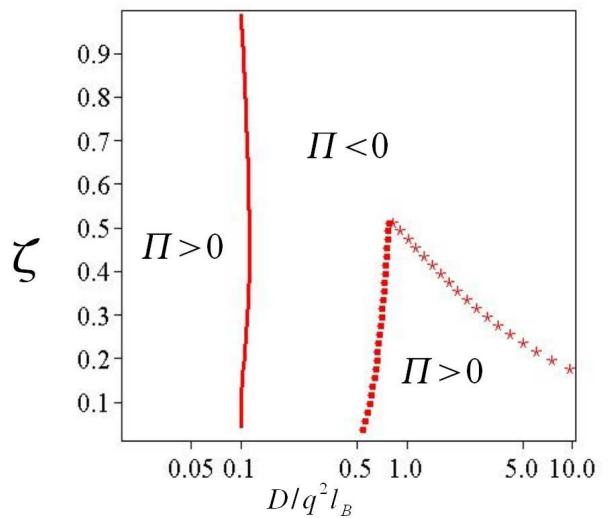}
 \end{center}
 \caption{Sketch of attractive and repulsive regimes, 
 as a function of the ratio $\zeta=\sigma_2/|\sigma_1|$ and the distance $D$ (in log scale). The repulsive island on the left hand side, delimitated by a continuous line, shows $D_1^*$ where the pressure in Eq. (\ref{eq12}) vanishes. 
 The other repulsive region on the right hand side, delimitated by a dotted line (star symbols), 
 shows $D^*_{MF}=|\mu_1^{\rm eff}-\mu_2^{\rm eff}|$, where the effective Gouy-Chapman lengths are given 
 by  Eqs.\eqref{eq18}. These data are displayed provided they satisfy the constraint 
 $D^*_{MF}>D_{\rm bound}=a_\perp^{(1)}+a_\perp^{(2)}$. Likewise, the left boundary for this 
 mean-field repulsive island has been taken to be $D_{\rm bound}$, that is shown with a dotted line
 (square symbols). 
 Here, we have taken $\Xi_2 = 10$.}
 \label{Fig5}
 \end{figure}

When the separation distance between plates 1 and 2 is decreased from infinity,
the first inter-plate weak interaction regime is expected to be of mean-field type,
so that the presumably large distance attraction may turn into repulsion at 
a distance $D^*_{MF}=|\mu_1^{\rm eff}-\mu_2^{\rm eff}|$.
In this picture, the distance is varied at constant effective Gouy-Chapman lengths $\mu_1^{\rm eff}$ and $\mu_2^{\rm eff}$ given by:
\begin{equation}
\mu_i^{\rm eff} \,=\, 1.92 \, \mu_i \, e^{0.78 \sqrt{\Xi_i}} \, , \ i=1,2.
\label{eq18}
\end{equation}
If $|\mu_1^{\rm eff}-\mu_2^{\rm eff}|$ is significantly larger than the characteristic 
thresholds obtained in the previous subsections, we should have the
following ``reentrant'' sequence \{attraction $\to$ repulsion $\to$ attraction $\to$ repulsion\}
as $D$ decreases. The first transition is described by a mean-field argument, and the
last one by strong-coupling considerations, but the intermediate transition
\{repulsion $\to$ attraction\} occurs in a crossover region that resists
our theoretical understanding, and where additional \{repulsion $\to$ attraction\} transitions might take place. 
A related question deals with the lower bound
for the distance $D_{\rm bound}$ below which the mean-field profiles are no longer accurate.
For the sake of completeness, we will consider below that $D_{\rm bound}=a_\perp^{(1)}+a_\perp^{(2)}$.
Depending on the respective surface charge densities $\sigma_1$ and $\sigma_2$, we can then 
discriminate between two distinct situations: 
\begin{itemize} 
 \item $|\mu_1^{\rm eff}-\mu_2^{\rm eff}| < D_{bound}$. The interaction between the two plates is always attractive at large distances (mean-field regime), and then at short separation distances, the 
 strong-coupling phenomenology described in the first part of the paper prevails.
 \item $|\mu_1^{\rm eff}-\mu_2^{\rm eff}| > D_{bound}$. There is then already a transition between attraction and repulsion in the mean-field regime. By decreasing further the distance $D$ 
 and entering the short distance limit, one should observe another attractive range, as expressed in Eq.\eqref{eq10} for instance, before repulsion sets in at even smaller separations.
\end{itemize}
More complicated scenarios could be envisioned, but we summarize in
Fig. \ref{Fig5} the simplest possible, and provide a phase diagram 
obtained when considering that the Bjerrum pairs do contribute to the pressure, as 
in section \ref{ssec:osmowith}. 
We note that, for the parameters chosen, there is a reentrant behaviour observed 
with respect to the separation distance, in a large fraction of the 
($\zeta,D$) plane, more specifically when $\zeta <0.5$ (this threshold depends on the
value of $\Xi_2$ chosen, and increases with $\Xi_2$). 
We remind that the bottom part of the diagram, more specifically for
$\zeta < 1/(2\Xi_2+1)$, corresponds to a region where our arguments do not apply,
as discussed in section \ref{ssec:osmowith}. In this region, our short scale analysis
provides an ``all repulsive'' behaviour, and we may then speculate that repulsion
persists up to the effective mean-field threshold indicated by the asterisks,
which corresponds to large distances, on the order of $100 \, q^2 l_B$ or more.

\section{Discussion and conclusion}
\label{sec:concl}

In this paper, we have analyzed the interaction of two oppositely charged parallel interfaces,
each neutralized by its own counter-ions, without other micro-ions involved
(salt-free case, but with two species of micro-ions, of opposite signs). 
We have shown that a repulsive behaviour, quite expectedly, is always present 
at short enough separations; it simply stems from the diverging entropy cost for confining
micro-ions in a slab of vanishing extension. Our analysis completes the known
Poisson-Boltzmann phenomenology by investigating the case of strong coulombic couplings.
Short distance expansions reveal that depending on the formation of Bjerrum
pairs between the oppositely charged micro-ions, an attractive regime may or may not
be observed. By formation of pairs, we understand here the wealth of different
self-assembly scenarios where the pairs may further associate into
more complex objects, such as chains or various crystals \cite{WL93,AML10}.
We did not attempt at a precise evaluation of the corresponding contribution
to the pressure --a particularly demanding task-- but rather, we analyzed
limiting cases where this unknown contribution is bounded by reasonable values,
see section \ref{ssec:with}.
We have supplemented our short distance analysis by a more speculative
investigation of the large distance behaviour, from which a  ``phase diagram''
was put forward, with reentrant features between attraction and repulsion 
as the distance $D$ between the plates is varied. The experimental observation 
of such features would imply that other sorts of interactions, such as van der Waals,
do not modify the main effects uncovered.

In our approach, single particle arguments play a crucial role and 
allow us to compute the density of micro-ions, from which the equation of state
follows. 
These single particle arguments, however, are {\it a priori} not restricted to
strongly coupled interfaces, but can equally be invoked when the coupling
parameters $\Xi_1$ and $\Xi_2$ are small (see e.g. section 3.3 of Ref \cite{Moreira02}
and in particular Fig. 16 for simulation results backing up this statement 
in the like-charge case $\sigma_1=\sigma_2$). 
Indeed, when $D$ becomes smaller 
than the characteristic lateral distance $a_\perp$ between ions, these ions
effectively decouple and feel the external potential of the plates only
(we are here concerned with the ionic density dependence on the $z$
coordinate, perpendicular to the plate; in the transverse direction,
parallel to the plate, a correlation hole remains around each particle, of typical
size $q^2 l_B$).
As a consequence, the pressures given by Eqs. (\ref{eq6}), (\ref{eq10})
and (\ref{eq12}) still hold with the same range of validity,
for $\Xi_i \to 0$. In the corresponding distance range, the Poisson-Boltzmann 
results break down due to discreteness effects
[it is therefore essential here to make a clear distinction between 
Poisson-Boltzmann theory, and the low $\Xi$ limit of the original
model dealing with discrete particles ; Poisson-Boltzmann approach considers
from the outset continuous density fields, and can therefore not
be expected to hold at separation distances such that discreteness effects
do matter, i.e. when $D< a_\perp$; the adequacy of Poisson-Boltzmann to describe
the low $\Xi$ physics should then be understood as a statement which excludes
a small range of short separations $D$]. From the analysis of section 
\ref{sec:2}, we learn that when 
Bjerrum pair formation can be neglected, the threshold distance $D^*$ where 
repulsive behaviour sets in is still given by the Poisson-Boltzmann result 
$|\mu_2-\mu_1|=\mu_2-\mu_1$, provided this length is larger than both characteristic 
distances $a_\perp^{(1)}$ and $a_\perp^{(2)}$. In the opposite case,
when $|\mu_2-\mu_1|< \hbox{inf}(a_\perp^{(1)},a_\perp^{(2)})$,
we may speculate that $D^*$ lies between $a_\perp^{(1)}$ and $a_\perp^{(2)}$,
since the single particle argument which holds at smaller separations
leads to repulsion, while the Poisson-Boltzmann theory yields attraction
at larger separations \cite{Dahirel}.
If on the other hand, Bjerrum pairs form and contribute to the pressure
through their mean density (see section \ref{ssec:osmowith}), 
we have seen that $D^* = \mu_2$, which is thus larger than the Poisson-Boltzmann
result $\mu_2-\mu_1$. However, this result only holds provided $\zeta> (1+ 2\Xi_2)^{-1} \simeq 1$
(we are still considering the low $\Xi_i$ limit).
Given that $\zeta \leq 1$ by definition (i.e. $\sigma_2 < |\sigma_1|$), 
we see that here, the single particle picture
does not apply up to $\mu_2$ (except in a small $\zeta$ region close 
to 1), which means that $D^*$ is larger than
$A_\perp = [\pi(|\sigma_1|-\sigma_2)]^{-1/2}$. It can be checked that
generically, this length is smaller than the Poisson-Boltzmann prediction 
$\mu_2-\mu_1$, except again in a small $\zeta$-region around unity. 

{ \it Acknowledgment. This work has been partly supported by the EPSRC grant $N^{\circ}$ EP/I000844/1.}

\bibliographystyle{apsrev}
\bibliography{biblio}

\begin{thebibliography}{44}
\expandafter\ifx\csname natexlab\endcsname\relax\def\natexlab#1{#1}\fi
\expandafter\ifx\csname bibnamefont\endcsname\relax
  \def\bibnamefont#1{#1}\fi
\expandafter\ifx\csname bibfnamefont\endcsname\relax
  \def\bibfnamefont#1{#1}\fi
\expandafter\ifx\csname citenamefont\endcsname\relax
  \def\citenamefont#1{#1}\fi
\expandafter\ifx\csname url\endcsname\relax
  \def\url#1{\texttt{#1}}\fi
\expandafter\ifx\csname urlprefix\endcsname\relax\def\urlprefix{URL }\fi
\providecommand{\bibinfo}[2]{#2}
\providecommand{\eprint}[2][]{\url{#2}}

\bibitem[{\citenamefont{Hansen and L\"owen}(2000)}]{HL00}
\bibinfo{author}{\bibfnamefont{J.-P.} \bibnamefont{Hansen}} \bibnamefont{and}
  \bibinfo{author}{\bibfnamefont{H.}~\bibnamefont{L\"owen}},
  \bibinfo{journal}{Annu. Rev. Phys. Chem.} \textbf{\bibinfo{volume}{51}},
  \bibinfo{pages}{209} (\bibinfo{year}{2000}).

\bibitem[{\citenamefont{Levin}(2002)}]{Levin}
\bibinfo{author}{\bibfnamefont{Y.}~\bibnamefont{Levin}}, \bibinfo{journal}{Rep.
  Prog. Phys.} \textbf{\bibinfo{volume}{65}}, \bibinfo{pages}{1577}
  (\bibinfo{year}{2002}).

\bibitem[{\citenamefont{Messina}(2009)}]{Messina09}
\bibinfo{author}{\bibfnamefont{R.}~\bibnamefont{Messina}}, \bibinfo{journal}{J.
  Phys.: Condens. Matter} \textbf{\bibinfo{volume}{21}},
  \bibinfo{pages}{113102} (\bibinfo{year}{2009}).

\bibitem[{\citenamefont{Ben-Tal}(1995)}]{Ben-Tal}
\bibinfo{author}{\bibfnamefont{N.}~\bibnamefont{Ben-Tal}}, \bibinfo{journal}{J.
  Phys. Chem.} \textbf{\bibinfo{volume}{99}}, \bibinfo{pages}{9642}
  (\bibinfo{year}{1995}).

\bibitem[{\citenamefont{Wu et~al.}(2000)\citenamefont{Wu, Bratko, Blanch, and
  Prausnitz}}]{Wu}
\bibinfo{author}{\bibfnamefont{J.~Z.} \bibnamefont{Wu}},
  \bibinfo{author}{\bibfnamefont{D.}~\bibnamefont{Bratko}},
  \bibinfo{author}{\bibfnamefont{H.~W.} \bibnamefont{Blanch}},
  \bibnamefont{and} \bibinfo{author}{\bibfnamefont{J.~M.}
  \bibnamefont{Prausnitz}}, \bibinfo{journal}{Phys. Rev. E}
  \textbf{\bibinfo{volume}{62}}, \bibinfo{pages}{5273} (\bibinfo{year}{2000}).

\bibitem[{\citenamefont{Tull et~al.}(2007)\citenamefont{Tull, Bartlett, and
  Ryan}}]{Tull}
\bibinfo{author}{\bibfnamefont{E.~J.} \bibnamefont{Tull}},
  \bibinfo{author}{\bibfnamefont{P.}~\bibnamefont{Bartlett}}, \bibnamefont{and}
  \bibinfo{author}{\bibfnamefont{K.~R.} \bibnamefont{Ryan}},
  \bibinfo{journal}{Langmuir} \textbf{\bibinfo{volume}{23}},
  \bibinfo{pages}{7859} (\bibinfo{year}{2007}).

\bibitem[{\citenamefont{Jens~Ryd\'en et~al.}(2005)\citenamefont{Jens~Ryd\'en,
  Ullner, and Linse}}]{Linse}
\bibinfo{author}{\bibfnamefont{J.}~\bibnamefont{Jens~Ryd\'en}},
  \bibinfo{author}{\bibfnamefont{M.}~\bibnamefont{Ullner}}, \bibnamefont{and}
  \bibinfo{author}{\bibfnamefont{P.}~\bibnamefont{Linse}}, \bibinfo{journal}{J.
  Chem. Phys.} \textbf{\bibinfo{volume}{123}}, \bibinfo{pages}{034909}
  (\bibinfo{year}{2005}).

\bibitem[{\citenamefont{Dahirel and Hansen}(2009)}]{Hansen}
\bibinfo{author}{\bibfnamefont{V.}~\bibnamefont{Dahirel}} \bibnamefont{and}
  \bibinfo{author}{\bibfnamefont{J.-P.} \bibnamefont{Hansen}},
  \bibinfo{journal}{J. Chem. Phys.} \textbf{\bibinfo{volume}{131}},
  \bibinfo{pages}{084902} (\bibinfo{year}{2009}).

\bibitem[{\citenamefont{Jonsson and Stahlberg}(1999)}]{Jonsson}
\bibinfo{author}{\bibfnamefont{B.}~\bibnamefont{Jonsson}} \bibnamefont{and}
  \bibinfo{author}{\bibfnamefont{J.}~\bibnamefont{Stahlberg}},
  \bibinfo{journal}{Col. Surf. B} \textbf{\bibinfo{volume}{14}},
  \bibinfo{pages}{67} (\bibinfo{year}{1999}).

\bibitem[{\citenamefont{Bigdeli et~al.}(2008)\citenamefont{Bigdeli, Talasaz,
  Stahl, Persson, Ronaghi, Davis, and Nemat-Gorgani}}]{Bigdeli}
\bibinfo{author}{\bibfnamefont{S.}~\bibnamefont{Bigdeli}},
  \bibinfo{author}{\bibfnamefont{A.~H.} \bibnamefont{Talasaz}},
  \bibinfo{author}{\bibfnamefont{P.}~\bibnamefont{Stahl}},
  \bibinfo{author}{\bibfnamefont{H.}~\bibnamefont{Persson}},
  \bibinfo{author}{\bibfnamefont{M.}~\bibnamefont{Ronaghi}},
  \bibinfo{author}{\bibfnamefont{R.~W.} \bibnamefont{Davis}}, \bibnamefont{and}
  \bibinfo{author}{\bibfnamefont{M.}~\bibnamefont{Nemat-Gorgani}},
  \bibinfo{journal}{Biotec. Bioen} \textbf{\bibinfo{volume}{100}},
  \bibinfo{pages}{19} (\bibinfo{year}{2008}).

\bibitem[{\citenamefont{Morones~{\it et al.}}(2005)}]{Morones}
\bibinfo{author}{\bibfnamefont{J.-R.} \bibnamefont{Morones~{\it et al.}}},
  \bibinfo{journal}{Nanotechnology} \textbf{\bibinfo{volume}{16}},
  \bibinfo{pages}{2346} (\bibinfo{year}{2005}).

\bibitem[{\citenamefont{Jones et~al.}(2003)\citenamefont{Jones, Shanahan,
  Berman, and Thornton}}]{Jones}
\bibinfo{author}{\bibfnamefont{S.}~\bibnamefont{Jones}},
  \bibinfo{author}{\bibfnamefont{H.}~\bibnamefont{Shanahan}},
  \bibinfo{author}{\bibfnamefont{H.}~\bibnamefont{Berman}}, \bibnamefont{and}
  \bibinfo{author}{\bibfnamefont{J.}~\bibnamefont{Thornton}},
  \bibinfo{journal}{Nucleic Acids Research} \textbf{\bibinfo{volume}{31}},
  \bibinfo{pages}{7189} (\bibinfo{year}{2003}).

\bibitem[{\citenamefont{Vervey and Overbeek}(1948)}]{DLVO}
\bibinfo{author}{\bibfnamefont{J.}~\bibnamefont{Vervey}} \bibnamefont{and}
  \bibinfo{author}{\bibfnamefont{J.~T.~G.} \bibnamefont{Overbeek}},
  \emph{\bibinfo{title}{Theory of the Stability of Lyophobic collo\"ids}}
  (\bibinfo{publisher}{Elsevier: Amsterdam}, \bibinfo{year}{1948}).

\bibitem[{\citenamefont{Bhattacharjee and Elimelech}(1997)}]{SEI}
\bibinfo{author}{\bibfnamefont{S.}~\bibnamefont{Bhattacharjee}}
  \bibnamefont{and}
  \bibinfo{author}{\bibfnamefont{M.}~\bibnamefont{Elimelech}},
  \bibinfo{journal}{J.Colloid Inteface Sci.} \textbf{\bibinfo{volume}{193}},
  \bibinfo{pages}{273} (\bibinfo{year}{1997}).

\bibitem[{\citenamefont{Parsegian and Gingell}(1972)}]{Parsegian}
\bibinfo{author}{\bibfnamefont{V.~A.} \bibnamefont{Parsegian}}
  \bibnamefont{and} \bibinfo{author}{\bibfnamefont{D.}~\bibnamefont{Gingell}},
  \bibinfo{journal}{Biophys. J.} \textbf{\bibinfo{volume}{12}},
  \bibinfo{pages}{1192} (\bibinfo{year}{1972}).

\bibitem[{\citenamefont{Ohshima}(1975)}]{Ohshima}
\bibinfo{author}{\bibfnamefont{H.}~\bibnamefont{Ohshima}},
  \bibinfo{journal}{Colloid and Polymer Sci.} \textbf{\bibinfo{volume}{253}},
  \bibinfo{pages}{150} (\bibinfo{year}{1975}).

\bibitem[{\citenamefont{Sens and Joanny}(2000)}]{Joanny}
\bibinfo{author}{\bibfnamefont{P.}~\bibnamefont{Sens}} \bibnamefont{and}
  \bibinfo{author}{\bibfnamefont{J.-F.} \bibnamefont{Joanny}},
  \bibinfo{journal}{Phys. Rev. Lett.} \textbf{\bibinfo{volume}{84}},
  \bibinfo{pages}{4862} (\bibinfo{year}{2000}).

\bibitem[{\citenamefont{Ben-Yaakov et~al.}(2007)\citenamefont{Ben-Yaakov,
  Burak, Andelman, and Safran}}]{Andelman07}
\bibinfo{author}{\bibfnamefont{D.}~\bibnamefont{Ben-Yaakov}},
  \bibinfo{author}{\bibfnamefont{Y.}~\bibnamefont{Burak}},
  \bibinfo{author}{\bibfnamefont{D.}~\bibnamefont{Andelman}}, \bibnamefont{and}
  \bibinfo{author}{\bibfnamefont{S.}~\bibnamefont{Safran}},
  \bibinfo{journal}{EPL} \textbf{\bibinfo{volume}{79}}, \bibinfo{pages}{48002}
  (\bibinfo{year}{2007}).

\bibitem[{\citenamefont{Paillusson et~al.}(2009)\citenamefont{Paillusson,
  Barbi, and Victor}}]{Pai09}
\bibinfo{author}{\bibfnamefont{F.}~\bibnamefont{Paillusson}},
  \bibinfo{author}{\bibfnamefont{M.}~\bibnamefont{Barbi}}, \bibnamefont{and}
  \bibinfo{author}{\bibfnamefont{J.-M.} \bibnamefont{Victor}},
  \bibinfo{journal}{Mol. Phys.} \textbf{\bibinfo{volume}{107}},
  \bibinfo{pages}{1379} (\bibinfo{year}{2009}).

\bibitem[{\citenamefont{Dahirel et~al.}(2009)\citenamefont{Dahirel, Paillusson,
  Jardat, Barbi, and Victor}}]{Dahirel}
\bibinfo{author}{\bibfnamefont{V.}~\bibnamefont{Dahirel}},
  \bibinfo{author}{\bibfnamefont{F.}~\bibnamefont{Paillusson}},
  \bibinfo{author}{\bibfnamefont{M.}~\bibnamefont{Jardat}},
  \bibinfo{author}{\bibfnamefont{M.}~\bibnamefont{Barbi}}, \bibnamefont{and}
  \bibinfo{author}{\bibfnamefont{J.-M.} \bibnamefont{Victor}},
  \bibinfo{journal}{Phys. Rev. Lett.} \textbf{\bibinfo{volume}{102}},
  \bibinfo{pages}{228101} (\bibinfo{year}{2009}).

\bibitem[{\citenamefont{Netz}(2001)}]{Netz01}
\bibinfo{author}{\bibfnamefont{R.}~\bibnamefont{Netz}}, \bibinfo{journal}{Eur.
  Phys. J. E} \textbf{\bibinfo{volume}{5}}, \bibinfo{pages}{557}
  (\bibinfo{year}{2001}).

\bibitem[{\citenamefont{Naji et~al.}(2005)\citenamefont{Naji, Jungblut,
  Moreira, and Netz}}]{Naji05}
\bibinfo{author}{\bibfnamefont{A.}~\bibnamefont{Naji}},
  \bibinfo{author}{\bibfnamefont{S.}~\bibnamefont{Jungblut}},
  \bibinfo{author}{\bibfnamefont{A.}~\bibnamefont{Moreira}}, \bibnamefont{and}
  \bibinfo{author}{\bibfnamefont{R.}~\bibnamefont{Netz}},
  \bibinfo{journal}{Physica A} \textbf{\bibinfo{volume}{352}},
  \bibinfo{pages}{131} (\bibinfo{year}{2005}).

\bibitem[{\citenamefont{Rouzina and Bloomfield}(1996)}]{Rouzina96}
\bibinfo{author}{\bibfnamefont{I.}~\bibnamefont{Rouzina}} \bibnamefont{and}
  \bibinfo{author}{\bibfnamefont{V.}~\bibnamefont{Bloomfield}},
  \bibinfo{journal}{J. Phys. Chem.} \textbf{\bibinfo{volume}{100}},
  \bibinfo{pages}{9977} (\bibinfo{year}{1996}).

\bibitem[{\citenamefont{Shklovskii}(1999)}]{Shklovskii99}
\bibinfo{author}{\bibfnamefont{B.~I.} \bibnamefont{Shklovskii}},
  \bibinfo{journal}{Phys. Rev. E} \textbf{\bibinfo{volume}{60}},
  \bibinfo{pages}{5802} (\bibinfo{year}{1999}).

\bibitem[{\citenamefont{Kandu\ifmmode~\check{c}\else \v{c}\fi{}
  et~al.}(2008)\citenamefont{Kandu\ifmmode~\check{c}\else \v{c}\fi{}, Trulsson,
  Naji, Burak, Forsman, and Podgornik}}]{Kanduc08}
\bibinfo{author}{\bibfnamefont{M.}~\bibnamefont{Kandu\ifmmode~\check{c}\else
  \v{c}\fi{}}}, \bibinfo{author}{\bibfnamefont{M.}~\bibnamefont{Trulsson}},
  \bibinfo{author}{\bibfnamefont{A.}~\bibnamefont{Naji}},
  \bibinfo{author}{\bibfnamefont{Y.}~\bibnamefont{Burak}},
  \bibinfo{author}{\bibfnamefont{J.}~\bibnamefont{Forsman}}, \bibnamefont{and}
  \bibinfo{author}{\bibfnamefont{R.}~\bibnamefont{Podgornik}},
  \bibinfo{journal}{Phys. Rev. E} \textbf{\bibinfo{volume}{78}},
  \bibinfo{pages}{061105} (\bibinfo{year}{2008}).

\bibitem[{\citenamefont{Jho et~al.}(2008)\citenamefont{Jho,
  Kandu\ifmmode~\check{c}\else \v{c}\fi{}, Naji, Podgornik, Kim, and
  Pincus}}]{Jho08}
\bibinfo{author}{\bibfnamefont{Y.~S.} \bibnamefont{Jho}},
  \bibinfo{author}{\bibfnamefont{M.}~\bibnamefont{Kandu\ifmmode~\check{c}\else
  \v{c}\fi{}}}, \bibinfo{author}{\bibfnamefont{A.}~\bibnamefont{Naji}},
  \bibinfo{author}{\bibfnamefont{R.}~\bibnamefont{Podgornik}},
  \bibinfo{author}{\bibfnamefont{M.~W.} \bibnamefont{Kim}}, \bibnamefont{and}
  \bibinfo{author}{\bibfnamefont{P.~A.} \bibnamefont{Pincus}},
  \bibinfo{journal}{Phys. Rev. Lett.} \textbf{\bibinfo{volume}{101}},
  \bibinfo{pages}{188101} (\bibinfo{year}{2008}).

\bibitem[{\citenamefont{Dean et~al.}(2009)\citenamefont{Dean, Horgan, Naji, and
  Podgornik}}]{Dean09}
\bibinfo{author}{\bibfnamefont{D.}~\bibnamefont{Dean}},
  \bibinfo{author}{\bibfnamefont{R.}~\bibnamefont{Horgan}},
  \bibinfo{author}{\bibfnamefont{A.}~\bibnamefont{Naji}}, \bibnamefont{and}
  \bibinfo{author}{\bibfnamefont{R.}~\bibnamefont{Podgornik}},
  \bibinfo{journal}{J. Chem. Phys.} \textbf{\bibinfo{volume}{130}},
  \bibinfo{pages}{094504} (\bibinfo{year}{2009}).

\bibitem[{\citenamefont{Kandu\v{c} et~al.}(2010)\citenamefont{Kandu\v{c}, Naji,
  Forsman, and Podgornik}}]{Kanduc10}
\bibinfo{author}{\bibfnamefont{M.}~\bibnamefont{Kandu\v{c}}},
  \bibinfo{author}{\bibfnamefont{A.}~\bibnamefont{Naji}},
  \bibinfo{author}{\bibfnamefont{J.}~\bibnamefont{Forsman}}, \bibnamefont{and}
  \bibinfo{author}{\bibfnamefont{R.}~\bibnamefont{Podgornik}},
  \bibinfo{journal}{J. Chem. Phys.} \textbf{\bibinfo{volume}{132}},
  \bibinfo{pages}{124701} (\bibinfo{year}{2010}).

\bibitem[{\citenamefont{Hatlo and Lue}(2010)}]{Hatlo10}
\bibinfo{author}{\bibfnamefont{M.}~\bibnamefont{Hatlo}} \bibnamefont{and}
  \bibinfo{author}{\bibfnamefont{L.}~\bibnamefont{Lue}}, \bibinfo{journal}{EPL}
  \textbf{\bibinfo{volume}{89}}, \bibinfo{pages}{25002} (\bibinfo{year}{2010}).

\bibitem[{\citenamefont{Kandu\v{c} et~al.}(2011)\citenamefont{Kandu\v{c}, Naji,
  Forsman, and Podgornik}}]{Kanduc11}
\bibinfo{author}{\bibfnamefont{M.}~\bibnamefont{Kandu\v{c}}},
  \bibinfo{author}{\bibfnamefont{A.}~\bibnamefont{Naji}},
  \bibinfo{author}{\bibfnamefont{J.}~\bibnamefont{Forsman}}, \bibnamefont{and}
  \bibinfo{author}{\bibfnamefont{R.}~\bibnamefont{Podgornik}},
  \bibinfo{journal}{arXiv:1101.1362}  (\bibinfo{year}{2011}).

\bibitem[{\citenamefont{\ifmmode~\check{S}\else \v{S}\fi{}amaj and
  Trizac}(2011{\natexlab{a}})}]{ST11}
\bibinfo{author}{\bibfnamefont{L.}~\bibnamefont{\ifmmode~\check{S}\else
  \v{S}\fi{}amaj}} \bibnamefont{and}
  \bibinfo{author}{\bibfnamefont{E.}~\bibnamefont{Trizac}},
  \bibinfo{journal}{Eur. Phys. J. E} \textbf{\bibinfo{volume}{34}},
  \bibinfo{pages}{20} (\bibinfo{year}{2011}{\natexlab{a}}).

\bibitem[{\citenamefont{\ifmmode~\check{S}\else \v{S}\fi{}amaj and
  Trizac}(2011{\natexlab{b}})}]{ST11prl}
\bibinfo{author}{\bibfnamefont{L.}~\bibnamefont{\ifmmode~\check{S}\else
  \v{S}\fi{}amaj}} \bibnamefont{and}
  \bibinfo{author}{\bibfnamefont{E.}~\bibnamefont{Trizac}},
  \bibinfo{journal}{Phys. Rev. Lett.} \textbf{\bibinfo{volume}{106}},
  \bibinfo{pages}{078301} (\bibinfo{year}{2011}{\natexlab{b}}).

\bibitem[{\citenamefont{Henderson and Blum}(1978)}]{Henderson}
\bibinfo{author}{\bibfnamefont{D.}~\bibnamefont{Henderson}} \bibnamefont{and}
  \bibinfo{author}{\bibfnamefont{L.}~\bibnamefont{Blum}}, \bibinfo{journal}{J.
  Chem. Phys.} \textbf{\bibinfo{volume}{69}}, \bibinfo{pages}{5441}
  (\bibinfo{year}{1978}).

\bibitem[{\citenamefont{Wennerstr\"om et~al.}(1982)\citenamefont{Wennerstr\"om,
  J\"onsson, and Linse}}]{Wennerstrom}
\bibinfo{author}{\bibfnamefont{H.}~\bibnamefont{Wennerstr\"om}},
  \bibinfo{author}{\bibfnamefont{B.}~\bibnamefont{J\"onsson}},
  \bibnamefont{and} \bibinfo{author}{\bibfnamefont{P.}~\bibnamefont{Linse}},
  \bibinfo{journal}{J. Chem. Phys.} \textbf{\bibinfo{volume}{76}},
  \bibinfo{pages}{4665} (\bibinfo{year}{1982}).

\bibitem[{\citenamefont{Zwanikken and van Roij}(2009)}]{Roij}
\bibinfo{author}{\bibfnamefont{J.}~\bibnamefont{Zwanikken}} \bibnamefont{and}
  \bibinfo{author}{\bibfnamefont{R.}~\bibnamefont{van Roij}},
  \bibinfo{journal}{J.Phys.: Condens. Matt.} \textbf{\bibinfo{volume}{21}},
  \bibinfo{pages}{4241102} (\bibinfo{year}{2009}).

\bibitem[{\citenamefont{Burak et~al.}(2004)\citenamefont{Burak, Andelman, and
  Orland}}]{Burak}
\bibinfo{author}{\bibfnamefont{Y.}~\bibnamefont{Burak}},
  \bibinfo{author}{\bibfnamefont{D.}~\bibnamefont{Andelman}}, \bibnamefont{and}
  \bibinfo{author}{\bibfnamefont{H.}~\bibnamefont{Orland}},
  \bibinfo{journal}{Phys. Rev. E} \textbf{\bibinfo{volume}{70}},
  \bibinfo{pages}{016102} (\bibinfo{year}{2004}).

\bibitem[{\citenamefont{Chen and Weeks}(2006)}]{Chen06}
\bibinfo{author}{\bibfnamefont{Y.-G.} \bibnamefont{Chen}} \bibnamefont{and}
  \bibinfo{author}{\bibfnamefont{J.}~\bibnamefont{Weeks}},
  \bibinfo{journal}{Proc. Nat. Acad. Sci. USA} \textbf{\bibinfo{volume}{103}},
  \bibinfo{pages}{7560} (\bibinfo{year}{2006}).

\bibitem[{\citenamefont{Santangelo}(2006)}]{Santangelo06}
\bibinfo{author}{\bibfnamefont{C.~D.} \bibnamefont{Santangelo}},
  \bibinfo{journal}{Phys. Rev. E} \textbf{\bibinfo{volume}{73}},
  \bibinfo{pages}{041512} (\bibinfo{year}{2006}).

\bibitem[{\citenamefont{Buyukdagli et~al.}(2010)\citenamefont{Buyukdagli,
  Manghi, and Palmeri}}]{Buyukdagli10}
\bibinfo{author}{\bibfnamefont{S.}~\bibnamefont{Buyukdagli}},
  \bibinfo{author}{\bibfnamefont{M.}~\bibnamefont{Manghi}}, \bibnamefont{and}
  \bibinfo{author}{\bibfnamefont{J.}~\bibnamefont{Palmeri}},
  \bibinfo{journal}{Phys. Rev. Lett.} \textbf{\bibinfo{volume}{105}},
  \bibinfo{pages}{158103} (\bibinfo{year}{2010}).

\bibitem[{\citenamefont{Boroudjerdi et~al.}(2005)\citenamefont{Boroudjerdi,
  Kim, Naji, Netz, Schlagberger, and Serr}}]{Boroudjerdi05}
\bibinfo{author}{\bibfnamefont{H.}~\bibnamefont{Boroudjerdi}},
  \bibinfo{author}{\bibfnamefont{Y.-W.} \bibnamefont{Kim}},
  \bibinfo{author}{\bibfnamefont{A.}~\bibnamefont{Naji}},
  \bibinfo{author}{\bibfnamefont{R.}~\bibnamefont{Netz}},
  \bibinfo{author}{\bibfnamefont{X.}~\bibnamefont{Schlagberger}},
  \bibnamefont{and} \bibinfo{author}{\bibfnamefont{A.}~\bibnamefont{Serr}},
  \bibinfo{journal}{Phys. Rep.} \textbf{\bibinfo{volume}{416}},
  \bibinfo{pages}{129} (\bibinfo{year}{2005}).

\bibitem[{\citenamefont{dos Santos et~al.}(2009)\citenamefont{dos Santos,
  Diehl, and Levin}}]{dosSantos}
\bibinfo{author}{\bibfnamefont{A.}~\bibnamefont{dos Santos}},
  \bibinfo{author}{\bibfnamefont{A.}~\bibnamefont{Diehl}}, \bibnamefont{and}
  \bibinfo{author}{\bibfnamefont{Y.}~\bibnamefont{Levin}}, \bibinfo{journal}{J.
  Chem. Phys.} \textbf{\bibinfo{volume}{130}}, \bibinfo{pages}{124110}
  (\bibinfo{year}{2009}).

\bibitem[{\citenamefont{Weis and Levesque}(1993)}]{WL93}
\bibinfo{author}{\bibfnamefont{J.~J.} \bibnamefont{Weis}} \bibnamefont{and}
  \bibinfo{author}{\bibfnamefont{D.}~\bibnamefont{Levesque}},
  \bibinfo{journal}{Phys. Rev. Lett.} \textbf{\bibinfo{volume}{71}},
  \bibinfo{pages}{2729} (\bibinfo{year}{1993}).

\bibitem[{\citenamefont{Assoud et~al.}(2010)\citenamefont{Assoud, Messina, and
  L\"owen}}]{AML10}
\bibinfo{author}{\bibfnamefont{L.}~\bibnamefont{Assoud}},
  \bibinfo{author}{\bibfnamefont{R.}~\bibnamefont{Messina}}, \bibnamefont{and}
  \bibinfo{author}{\bibfnamefont{H.}~\bibnamefont{L\"owen}},
  \bibinfo{journal}{EPL} \textbf{\bibinfo{volume}{89}}, \bibinfo{pages}{36001}
  (\bibinfo{year}{2010}).

\bibitem[{\citenamefont{Moreira and Netz}(2002)}]{Moreira02}
\bibinfo{author}{\bibfnamefont{A.}~\bibnamefont{Moreira}} \bibnamefont{and}
  \bibinfo{author}{\bibfnamefont{R.}~\bibnamefont{Netz}},
  \bibinfo{journal}{Eur. Phys. J. E} \textbf{\bibinfo{volume}{8}},
  \bibinfo{pages}{33} (\bibinfo{year}{2002}).

\end{thebibliography}

\end{document}